\documentclass{emulateapj}

\shorttitle{3C 279 Flare with Fermi-II Acceleration}
\shortauthors{Asano \& Hayashida}

\begin{document}
\pagenumbering{arabic}
\title{
The Most Intensive Gamma-Ray Flare of Quasar 3C 279
with the Second-Order Fermi Acceleration
}
\author{\scshape Katsuaki Asano,
and
Masaaki Hayashida}
\email{asanok@icrr.u-tokyo.ac.jp, mahaya@icrr.u-tokyo.ac.jp}

\affil{Institute for Cosmic Ray Research, The University of Tokyo,
5-1-5 Kashiwanoha, Kashiwa, Chiba 277-8582, Japan}

\date{Submitted; accepted}

\begin{abstract}

The very short and bright flare of 3C 279 detected with {\it Fermi}-LAT
in 2013 December
is tested by a model with stochastic electron acceleration
by turbulences.
Our time-dependent simulation shows that the very hard spectrum and
asymmetric light curve
are successfully reproduced by changing only the magnetic field
from the value in the steady period.
The maximum energy of electrons drastically grows
with the decrease of the magnetic field, which yields a hard photon spectrum
as observed.
Rapid cooling due to the inverse-Compton scattering
with the external photons reproduces the decaying feature of the light curve.
The inferred energy density of the magnetic field is much less
than the electron and photon energy densities.
The low magnetic field and short variability timescale
are unfavorable for the jet acceleration model
from the gradual Poynting flux dissipation.

\end{abstract}


\maketitle

\section{Introduction}
\label{sec:intro}

Multi-wavelength light curves of blazar flares show
complex and diversified features.
While in some cases there is a time lag between gamma-ray and X-ray/optical
flares \citep[e.g.][]{bla05,fos08,abd10,hay12},
in other cases an orphan flare in a certain wavelength
was detected \citep[e.g.][]{kra04,abd10b}.
Even if a time-dependent model is adopted,
such a variety of behaviors may be difficult to reproduce with
by a one-zone model \citep{kus00,kra02,asa14}.
While spatial gradients of the physical parameters in the emission regions
\citep{jan12} may explain some fraction of the lags,
some flares have spectral evolutions thatare too complex to be
modeled, even with time-dependent multi-zone radiative transfer simulations
\citep{time-dep}.
This may imply that inhomogeneous emission regions evolve with a longer timescale
than a dynamical one.
Such nontrivial properties in a blazar flare make it difficult
to probe physical processes such as electron acceleration
or cooling.

In 2013 December, the {\it Fermi}-Large Area Telescope (LAT)
detected one of the most intense flares
in the gamma-ray band from flat spectrum radio quasar (FSRQ) 3C~279,
reaching $\sim 1\times10^{-5}\,{\rm ph}\,{\rm cm}^{-2}\,{\rm s}^{-1}$
for the integrated flux above 100\,MeV~\citep[][hereafter H15]{hay15}.
The flux level is comparable to the historical maximum of
this source observed at the gamma-ray band~\citep{Weh98}.
The gamma-ray flare showed a very rapid variability
with an asymmetric time profile with a shorter rising time of
$\sim2$\,hr and a longer falling time of $\sim7$\,hr.
We can expect that this extraordinary flare was emitted from a sufficiently compact region
that can be regarded as homogeneous, which is different from other usual flares.
In this case, the decaying timescale may directly correspond to the cooling timescale,
and the flare is an ideal target for discussing the physical processes.

Another important property of the flare event of 3C 279, a very hard photon index of
$\Gamma_\gamma = 1.7\pm0.1$, was observed in the $>100$\,MeV band by {\it Fermi}-LAT.
Such a hard photon index has been rarely observed in FSRQs,
whose luminosity peak from inverse-Compton (IC)
scattering is usually located below 100\,MeV.
While the mean of the $\Gamma_\gamma$ distribution in FSRQs
corresponds to about 2.4 \citep{3LAC},
hard photon indices $\Gamma_\gamma < 2$ only have been occasionally observed in some bright
FSRQs during rapid flaring events~\citep{Pac14}.
In order to reproduce the hard photon index by IC scattering
in the fast cooling regime, 
the index of parent electrons should be much harder than two,
which can hardly be generated in a normal shock acceleration process.
In addition, the flare event of 3C~279 indicates a high Compton dominance parameter
$L_\gamma/L_{\rm syn} > 300$, leading to extremely low jet magnetization with
$L_{\rm B}/L_{\rm j} \lesssim 10^{-4}$~(H15).

To explain the flare event of 3C~279,
rather than assuming prompt electron injection by the shock acceleration,
we propose the stochastic acceleration (SA) model,
which is phenomenologically equivalent to the second-order Fermi acceleration
\citep[Fermi-II; e.g.][and references therein]{kat06,sta08,lef11}.
The SA may be driven
by magnetic reconnection \citep{laz12}.
Otherwise, hydrodynamical turbulences
that drive the acceleration are possibly induced via
the Kelvin--Helmholtz instability as an axial mode
\citep[e.g.][]{miz07},
or the Rayleigh--Taylor and Richtmyer--Meshkov instabilities
as radial modes \citep{mat13}.
Broadband spectra of blazars in the steady state
have been successfully fitted with recent SA models
\citep{asa14,dil14,kak15}.
The flare state should be also tested with such models
to show the wide-range applicability of the SA.
This is the first attempt to apply a Fermi-II model to
explain both broadband spectra and light curves of FSRQs simultaneously.

In this Letter, we perform time-dependent simulations
of the emissions from 3C 279 with the SA.
Starting from modeling a steady emission,
the gamma-ray flare is reproduced by decreasing the magnetic field
for the steady model.
We demonstrate that the SA model agrees
with the observed spectrum and light curve.

\section{Numerical Methods}
\label{sec:model}

We adopt the numerical simulation code used in \citet{asa14}.
In this model, a conical outflow with an opening angle
$\theta_{\rm j}=1/\Gamma$, where $\Gamma$ is its bulk Lorentz factor,
is ejected at radius $R=R_0$ from the central engine.
The evolutions of electron and photon energy distributions
are calculated in the comoving frame taking into account
the SA, synchrotron emission,
IC scattering with the Klein--Nishina effect,
adiabatic cooling, $\gamma \gamma$ pair production,
synchrotron self-absorption, and photon escape.
The SA is characterized by the energy diffusion
coefficient, $D(\varepsilon_{\rm e})=K \varepsilon_{\rm e}^q$.
\citet{asa14} conservatively assumed the Kolmogorov-like
diffusion as $q=5/3$.
Here, we adopt, however, $q=2$, which corresponds to
the hard-sphere scattering.
This choice leads to reasonable
spectra of 3C 279 without complicated assumptions such as
nontrivial temporal evolution of the diffusion coefficient
or electron injection rate.
If the cascade of the turbulence stops at a certain length scale
larger than the gyro-radius of the highest-energy electrons,
the mean free path of electrons becomes comparable to this scale
independently of electron energies.
In this case, the energy diffusion can be approximated as the hard-sphere
scattering \citep[e.g.][]{ber11}.
The recent magnetohydrodynamical simulations accompanying the inverse cascade
shows the $k^{-2}$ spectrum in turbulences \citep{zra14,bra15}, which also support
the hard-sphere approximation.

The volume we consider is a conical shell with a constant width of $W'=R_0/\Gamma$,
then the isotropically equivalent volume $V' = 4 \pi R^2 W'$
(the actual volume is $\pi \theta_{\rm j}^2 R^2 W'$).
Hereafter, the values in the shell frame are denoted with prime characters.
During the dynamical timescale $W'/c$ in the plasma frame,
electrons are injected at a constant rate $\dot{N}'_{\rm e}$ in the volume
$V'$ monoenergetically ($\gamma_{\rm e}=10$)
and accelerated with a constant coefficient $K'$.
As done in \citet{asa14}, we can consider the temporal evolutions
of the injection rate and the diffusion coefficient.
However, this simple model with constant $\dot{N}'_{\rm e}$ and $K'$
is sufficient to reproduce the photon spectrum of 3C 279.
The average magnetic field in the comoving frame is assumed to behave as
$B'=B_0 (R/R_0)^{-1}$.
The evolution of the photon spectrum for observers
is computed taking into account the relativistic motion and
curvature of the jet surface.

\section{Steady Model for the Active Period in 2009}
\label{sec:steady}

As a reference case, we consider
one of the most active periods in the gamma-ray band
during the first two years of the {\it Fermi}-LAT observations.
In \citet{hay12}, this period is denoted as period ``D'' in 2009.
Although this period corresponds to
an event with a prominent flare, the broadband spectrum in the paper
is averaged over five days.
If the emission zone is inside the broad emission region
as suggested by the short variability timescale
reported in H15,
the flare state is significantly longer than the variability
timescale.
Therefore, we adopt a steady emission model for period D
in 2009.

By assuming continuous steady ejection of the shells from $R=R_0$,
we model the steady photon spectrum, though the plasma and its emission
evolves with $R$ in the shell frame.
The model parameters are $R_0=0.023$ pc, $\Gamma=15$,
$K'=9 \times 10^{-6}~\mbox{s}^{-1}$ ($t_{\rm acc}=1/(2K')=0.35 W'/c$),
$\dot{N}'_{\rm e}=7.8 \times 10^{49}~\mbox{s}^{-1}$
($\dot{n}'_{\rm e}=0.26 (R/R_0)^{-2}~\mbox{cm}^{-3}~\mbox{s}^{-1}$),
and $B_0=7$ G. We adopt the same model
as that of \citet{hay12}
for the external radiation of the broad emission lines
with the photon temperature $T'_{\rm UV}=10 \Gamma$~eV
and the energy density $U'_{\rm UV}=8(\Gamma/15)^2~\mbox{erg}~\mbox{cm}^{-3}$
in the shell frame.
The cooling timescale in this external radiation field is
written as
\begin{eqnarray}
t_{\rm cool}=\frac{3 m_{\rm e} c}{4 \sigma_{\rm T} \gamma_{\rm e} U'_{\rm UV}}
=0.24 \frac{W'}{c} \left( \frac{\gamma_{\rm e}}{100} \right)^{-1}
\sim 0.7 t_{\rm acc} \left( \frac{\gamma_{\rm e}}{100} \right)^{-1}.
\end{eqnarray}

\begin{figure*}[t]
\centering
\epsscale{1.1}
\plotone{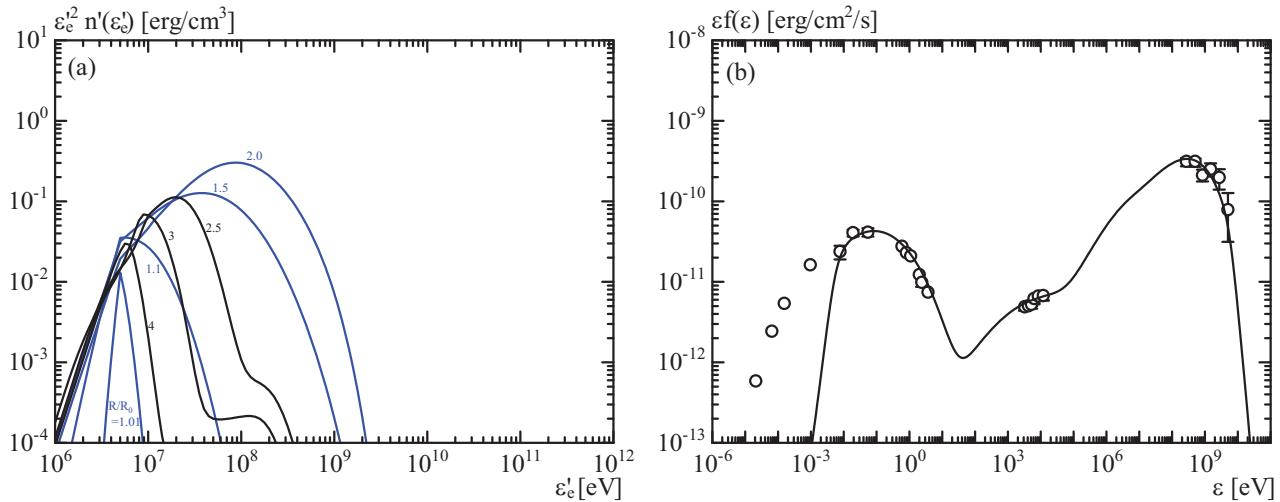}
\caption{(a) Evolution of the electron energy distribution
with increasing distance $R$
in the steady model. The numbers beside each line denote
$R/R_0$. The electron spectra during the acceleration phase
($R \leq 2 R_0$) are plotted with blue lines,
while those after the shutdown of the acceleration
($R>2R_0$) are black.
(b) Photon spectrum for the active period in 2009.
The open circles are measured flux points in period D
\citep[adapted from][]{hay12}.
The solid line is the model spectrum that is a superposition
of the emissions from the all shells of $R>R_0$.
\label{fig:ele1}}
\end{figure*}

As shown in Figure \ref{fig:ele1} (a),
electrons are continuously accelerated between $R=R_0$ and $2 R_0$,
then they are rapidly cooled via IC scattering
after the shutdown of the acceleration.
The electron spectrum at $2 R_0$ in the low-energy region
is consistent with the assumed power-law index of $p=1$
in the broken power-law model of \citet{hay12}.
In the highest-energy region, though the Klein--Nishina effect
suppresses the IC cooling effect, the synchrotron cooling
($U'_B = 1.9 (R/R_0)^{-2}~\mbox{erg}~\mbox{cm}^{-3}$) prevent
the acceleration above $100$ MeV.
The resultant photon spectrum well
reproduces the observed spectrum
from far-infrared to gamma-ray bands (see Figure \ref{fig:ele1} (b)).
The low-energy cutoff at $\sim 0.02$ eV
is due to the synchrotron self-absorption.
The X-ray flux is originated from the synchrotron self-Compton (SSC)
emission.
Those X-ray data strongly constrain the emission radius $R_0$.

Thus, our SA model can naturally produce a hard electron
spectrum, and the steady photon spectrum agrees with
the observed one in 2009.
Based on this result, we will probe the intensive flare in 2013
in the next section.

\section{Flare Model in 2013}
\label{sec:flare}

The most intensive flare denoted as period ``B'' (on MJD 56646)
in H15
shows a very hard spectrum with $\Gamma_\gamma \sim 1.7$
and a short variability with an hourly scale in the gamma-ray band
observed with {\it Fermi}-LAT.
During a short time interval of the gamma-ray flare period (0.2 days),
there were simultaneous optical observations,
whose results did not show any correlated variability with
the gamma-ray flare as presented in H15.
The X-ray observations in the period are available
from {\it Swift}-BAT transient monitor results
\footnote{http://swift.gsfc.nasa.gov/results/transients/weak/3C279/}
by the {\it Swift}-BAT team \citep{kri13}.
The data provided an upper limit in the 15-50 keV band.
We adopt the same values for $R_0$, $\Gamma$,
$T'_{\rm UV}$, and $U'_{\rm UV}$ as those in the previous section.
By changing $K'$, $\dot{N}'_{\rm e}$, and $B_0$,
we attempt to fit the spectrum of the flaring period B in 2013.

We consider one shell that contributes to the flare emission.
The energy diffusion coefficient and injection rate are slightly increased 
from the values in the steady model to
$K'=1.3 \times 10^{-5}~\mbox{s}^{-1}$ ($t_{\rm acc}=1/(2K')=0.25 W'/c$),
and $\dot{N}'_{\rm e}=2.5 \times 10^{50}~\mbox{s}^{-1}$
($\dot{n}'_{\rm e}=0.85 (R/R_0)^{-2}~\mbox{cm}^{-3}~\mbox{s}^{-1}$),
respectively.
Hereafter, we call this the ``fiducial'' flare model.
No significant concurrent flare in the optical bands implies
that the optical photons are emitted from other steady components.
As discussed in Section 4.3 in H15,
the lack of overall correlation between the optical and gamma-ray bands
in 2013-2014 also suggests the different origin 
of the optical component.
The synchrotron flux of the flare should be below
the observed flux level so that an upper limit for the magnetic field
in the flare zone will be given.
Here, we adopt a very low value of $B_0=0.25$ G.

\begin{figure*}[t]
\centering
\epsscale{1.1}
\plotone{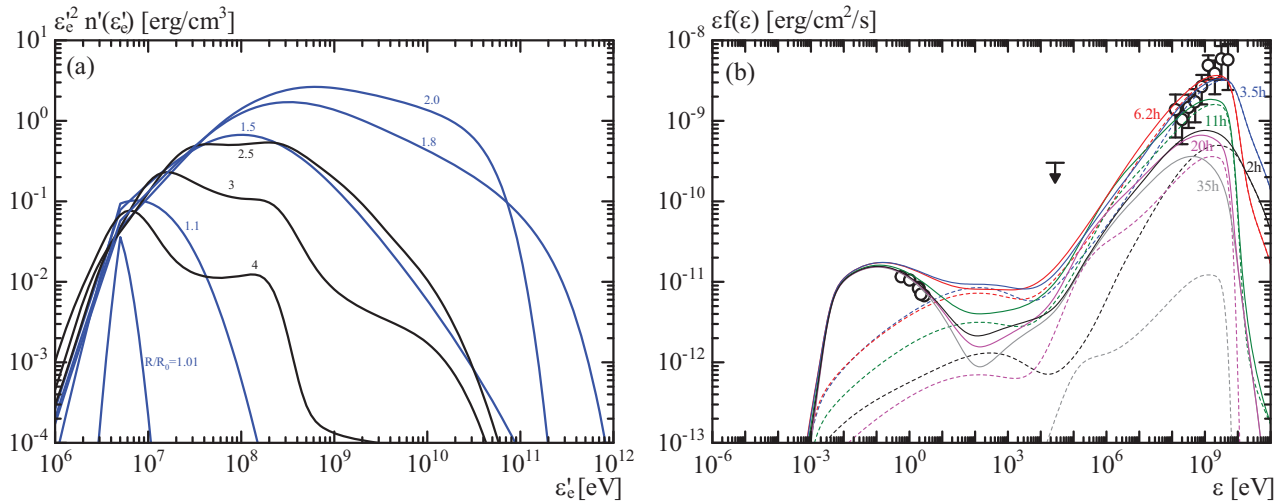}
\caption{(a) Evolution of the electron energy distribution
with increasing distance $R$
in the fiducial flare model.
The figure format is the same as that of Fig. 1(a).
(b) Photon spectrum for the most intense flare in 2013.
The open circles are measured flux points in period B
(adapted from H15), and the $2\sigma$
upper limit for the hard X-ray
is obtained from {\it Swift}-BAT data.
The solid lines are the model spectra at observation times
of 2 (black), 3.5 (blue), 6.2 (red), 11 (green), 20 (purple),
and 35 (gray) hours.
The model parameters for an underlying component,
which is consistent with the gamma-ray flux the day before the flare event
($0.19\times 10^{-5}$ photons~cm$^{-2}~\mbox{s}^{-1}$
above 0.1 GeV) and the optical data, are the same as those of the steady model
in \S \ref{sec:steady}
except for $B_0=3.8$ G and $\dot{N}'_{\rm e}=7.3 \times 10^{49}~\mbox{s}^{-1}$.
The dashed lines show the flare components only.
The observation time $t_{\rm obs}$
is measured from the first arrival time
of the photons escaped from the shell at $R=R_0$.
Photons are supposed to be emitted from a single shell
moving toward us.
\label{fig:ele2}}
\end{figure*}

Figure \ref{fig:ele2} (a) shows the evolution
of the electron energy distribution in this flare model.
Electrons are accelerated to higher energies compared to
the case in the steady model.
The power-law distributions above $10^9$~eV
are due to not only the larger $K'$ but also
the inefficiency of the synchrotron cooling
owing to the low magnetic field.
The secondary bumps at $\sim 2 \times 10^8$~eV for $R=3$-$4 R_0$
are attributed to the generation of secondary electron--positron pairs
via internal $\gamma \gamma$ absorption.

The resultant gamma-ray spectra shown in Figure \ref{fig:ele2} (b)
agree well with the observed gamma-ray data.
Here, we add an underlying component
(that overlaps the solid gray line in the figure)
to explain the gamma-ray flux before the flare and the optical data.
The flare spectrum has a higher synchrotron peak energy
than the model in H15, since our flare model
shows a drastic growth of the maximum energy of electrons
compared to the steady model.
As remarked above, the weak magnetic field strikingly
increases the maximum energy of electrons.
Even for this low magnetic field,
the optical flux is slightly enhanced during the flare.
The steady behavior of the optical light curve
may prefer a weaker magnetic field,
but we regard this as a conservative upper limit.

The sharp cutoff at $\sim 10^{10}$~eV in the photon spectrum
is due to the
$\gamma \gamma$ absorption inside the emission region.
Some fraction of photons above $10^{10}$~eV escape
from the shell.
The model flux at the $100$ GeV band
is still higher than the detection limit for \v{C}erenkov telescopes.
However, it should be noted that we have neglected the $\gamma \gamma$ absorption
after the escape from the shell.
The absorption by the broad emission lines during propagation
may greatly suppress the flux around $100$ GeV.

\begin{figure}[htb!]
\centering
\epsscale{1.0}
\plotone{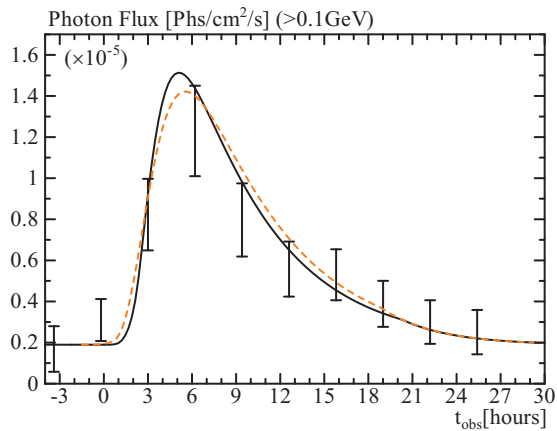}
\caption{Gamma-ray light curve of period B in 2013
with the flare model (solid black),
and ``High-$K$'' model (dashed orange).
The original data were obtained from H15.
In the model light curves,
the underlying gamma-ray level is the same as in Fig. \ref{fig:ele2}
($0.19\times 10^{-5}$ photons cm$^{-2}~\mbox{s}^{-1}$).
\label{fig:lc}}
\end{figure}

\begin{figure}[htb!]
\centering
\epsscale{1.0}
\plotone{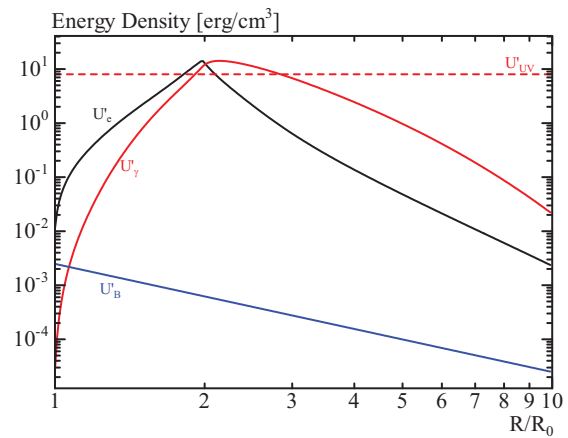}
\caption{Evolution of the energy densities in the shell frame
for the flare model.
The solid black, blue, and red lines show the value of
electrons, magnetic fields, and photons produced in the shell,
respectively. The red dashed line denotes the value of the external photons.
\label{fig:ened}}
\end{figure}

Even for the same values of $R_0$ and $\Gamma$
as in the steady model,
the light curve is well reproduced as shown in Figure \ref{fig:lc}.
Thus, the emission zones of the intense flare in 2013 and the active period in 2009
may be located at similar distances from the central engine.
The observed asymmetric profile in the light curve
is favorable for our simple one-shell emission-zone model.
The strong cooling due to the external IC yields
the rapid decay of the light curve.
The evolutions of energy densities in Figure \ref{fig:ened}
clearly show the energy input by the SA
and rapid cooling just after the end of the acceleration.
At $R=2 R_0$, the energy density ratio
of the magnetic field to electrons
is quite low as $U'_B/U'_{\rm e} \sim 4 \times 10^{-5}$.

\begin{figure}[htb!]
\centering
\epsscale{1.0}
\plotone{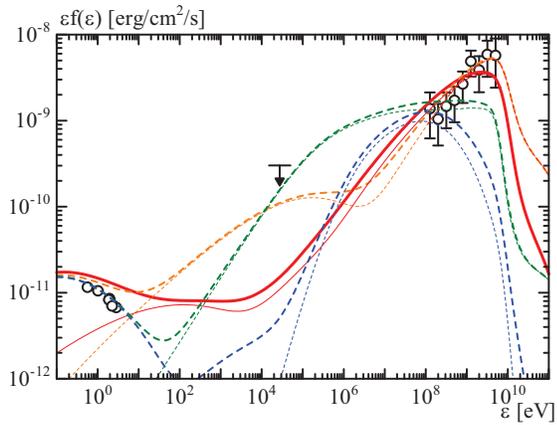}
\caption{Comparison of the model spectra at $t_{\rm obs}=6.2$~hr:
the same model as in Fig. \ref{fig:ele2} (solid red),
``High-$K$'' model (dashed orange), ``Low-$K$'' model (dashed blue),
and ``High-$\Gamma$'' model (dashed green).
The steady underlying component
is included as in Figure \ref{fig:ele2}.
The thin lines show the flare component only.
\label{fig:comp}}
\end{figure}

The observational constraints, of course,
do not determine the model parameter uniquely.
However, the essential parameter for determining the gamma-ray spectral shape
is only $K'$ in our model
(the role of $\dot{N}'_{\rm e}$ is just normalizing the flux level,
and the value of $B_0$ does not affect the gamma-ray spectral shape).
In Figure \ref{fig:comp}, we compare several photon spectral models
derived with different parameter sets.
When we reduce $K'$ by a factor of two from the fiducial model
(``Low-$K$'' model: $K' \to K' \times 0.5$, $\dot{N}'_{\rm e} \to
\dot{N}'_{\rm e} \times 4.8$, the others are the same),
the peak photon energy does not reach 10 GeV.
Conversely, we increase the diffusion coefficient
as shown in the ``High-$K$'' model
($K' \to K' \times 1.5$, $\dot{N}'_{\rm e} \to
\dot{N}'_{\rm e} \times 0.038$, $B_0 \to B_0 \times 0.4$, the others are the same).
In this case, we need an even weaker magnetic field.
The peak time of the light curve is delayed due to the lower $\dot{N}'_{\rm e}$
compared to the fiducial model.
In Figure \ref{fig:lc}, we shift
the light curve by 1.5 hr earlier.
Other physical parameters ($\Gamma$, etc.) of the jet were derived from the steady model.
However, as H15 supposed, we also try to increase $\Gamma$ in our model.
The initial radius should be increased as $\propto \Gamma^2$
to keep the variability timescale.
Such an example (``High-$\Gamma$'' model: $\Gamma \to \Gamma \times 2$,
$R_0 \to R_0 \times 4$,
$\dot{N}'_{\rm e} \to \dot{N}'_{\rm e} \times 0.039$,
$B_0 \to B_0 \times 0.16$, $K'$ is the same)
is shown in Figure \ref{fig:comp}.
Due to the relatively short $t_{\rm acc}$
($=0.18W'/c$), the maximum electron energy
grows as high as $\varepsilon'_{\rm e} \sim 10^{13}$~eV.
A very low magnetic field ($U'_B/U'_{\rm e} \lesssim 6 \times 10^{-5}$) is necessary again 
to suppress the synchrotron flux in the X-ray band.
The strong cooling due to the higher $U'_{\rm UV} \propto \Gamma^2$
makes the GeV spectrum too soft compared to the observed gamma-ray spectrum.

\section{Discussion}
\label{sec:sum}

The simple SA model can reasonably
explain the very hard spectrum and short variability
in the intensive flare in 2013.
The turbulence driving the particle acceleration
may be generated by the hydrodynamical instability
or the magnetic reconnection.

Compared to the steady model for the active period in 2009,
the drastic alteration we need
is the decrease of the magnetic field.
The other parameters have almost similar values.
The absence of the optical flare implies the weak magnetic field
($<0.25$~G).
The requirement of the magnetic field decrease at gamma-ray flare stages
was suggested by \citet{asa14} as well.
The required low magnetic field seems irrelevant to the energy source
for the particle acceleration.
Therefore, a hydrodynamical instability is responsible
for driving the SA.

When $\Gamma=15$, the variability timescale is consistent with $R_0=0.02$~pc
as shown in Figure \ref{fig:lc}.
This distance from the engine also agrees with the constraint
by the X-ray SSC component in the active period in 2009.
The size of the central engine may be
$\sim 3 r_{\rm g} \sim 8.6 \times 10^{-5}$~pc
for the black hole mass of $3 \times 10^8 M_\sun$.
If we adopt the simplest model for the jet acceleration
due to the magnetic energy dissipation \citep{dre02},
the bulk Lorentz factor at $R=R_0$ should be $<(R_0/3 r_{\rm g})^{1/3} \sim 6$,
which is inconsistent with the postulated value of $\Gamma$.
Given the variability timescale $\Delta t$,
the initial radius can be scaled as $R_0 \propto \Delta t \Gamma^2$
when we change $\Gamma$.
However, even in this case, the maximum Lorentz factor at $R_0$
inferred from the magnetic dissipation model 
increases by a factor of only $R_0^{1/3} \propto \Gamma^{2/3}$.
For the jet acceleration model by the Poynting flux dissipation,
not only the low magnetic field but also the short variability timescale
are problematic.
This problem is also raised for the very short gamma-ray flare
(a few hundreds seconds) of BL Lac objects like PKS 2155--304
\citep{aha07}. For such SSC-dominant objects, however,
the distance from the engine is not well constrained compared to FSRQs.

The tiny change of the diffusion coefficient $K'$,
in spite of the drastic decrease of the magnetic field,
seems enigmatic.
The assumed value of $q=2$ may be favorable for this invariant behavior of $K'$.
In this low magnetic field case,
the average energy gain per scattering may be proportional to $\beta_{\rm W}^2$,
where $\beta_{\rm W}$ is the average turbulence velocity,
rather than the Alfv\'en velocity.
The pitch angle diffusion approximation \citep{bla87}
and power-law magnetic turbulence of $\delta B^2 (k)=\delta B_0^2 k^{-q}$,
where $k$ is the wavenumber, leads to $K \propto \beta_{\rm W}^2 \delta B_0^2 B^{-q}$.
The last factor of $B^{-q}$ implies that electrons interact with
higher (lower) amplitude turbulence at longer (shorter)
wavelengths for a weaker (stronger) magnetic field.
If $\delta B_0 \propto B$, $q=2$ results in $K \propto \beta_{\rm W}^2$,
which is independent of $B$.
Alternatively, magnetic bottles as ``hard spheres'' \citep{ber11}
may be formed in turbulence independently of the strength of the magnetic field.
Such requirements for the
turbulence property motivate us to probe the hydrodynamical
instabilities in blazar jets.

\begin{acknowledgments}
The authors thank the anonymous referee for the useful comments.
We also thank F. Takahara, M. Kusunose, K. Toma,
J. Kakuwa, K. Nalewajko and G. M. Madejski for useful discussion.
This study is partially supported by Grants-in-Aid for Scientific Research
No.80399279 from the Ministry of Education,
Culture, Sports, Science and Technology (MEXT) of Japan.
\end{acknowledgments}

\end{document}